\documentclass[twocolumn,superscriptaddress,floatfix,preprintnumbers,prd, nofootinbib,hyperref]{revtex4-2}

\usepackage{amsmath,amssymb}
\begin{document}

\title{Constraints and Consistency of Rotating Regular Black Hole Thermodynamics}

\author{Qi-Quan Li}
\affiliation{School of Physics Science and Technology, Xinjiang University, Urumqi 830046, China}
\author{Yu Zhang}
\email[]{Contact author: zhangyu\_128@126.com}
\affiliation{Faculty of Science, Kunming University of Science and Technology, Kunming 650500, China}

\author{Hoernisa Iminniyaz}
\email[]{Contact author: wrns@xju.edu.cn}
\affiliation{School of Physics Science and Technology, Xinjiang University, Urumqi 830046, China}


\begin{abstract}
The thermodynamics of regular black holes has long suffered from a self-consistency problem: the Hawking temperature derived from the first law disagrees with the geometrically defined one, signaling that the thermodynamic quantities obtained in previous studies are unreliable. To resolve this, we construct a rotating ``mother'' black hole whose parameters are initially independent, forming an extended thermodynamic phase space. Thermodynamic quantities are first derived in this unconstrained phase space, and only then is the regularity condition imposed. We find a fundamental asymmetry between the geometric and thermodynamic derivations: the geometric temperature is invariant under the order of constraint application, whereas the thermodynamic temperature requires the correct order of operations to yield a physically meaningful result. This establishes a self-consistent thermodynamic framework for rotating regular black holes, in which all thermodynamic quantities are guaranteed to be physically correct, laying a rigorous foundation for investigating phase transitions and thermodynamic stability.

\end{abstract}

\maketitle

{\it Introduction} Black holes are one of the most famous predictions of general relativity, harboring singularities at their cores, where curvature diverges and physical laws break down. The singularity theorem proposed by Hawking and Penrose in 1970 states that a spacetime satisfying four specific physical conditions must be causally geodesically incomplete, which implies the existence of a singularity \cite{Hawking:1970zqf}. The problem of black hole singularities stems from general relativity, but general relativity itself breaks down at the singularity, exposing the incompleteness of current gravitational theory. Although it is believed that a quantum gravity theory is ultimately required to fully resolve the black hole singularity problem, such a theory has not yet been established. Therefore, attempts to address the singularity problem using semiclassical theories become particularly valuable.

The first regular black hole solution was proposed by Bardeen in 1968 \cite{bardeen1968non}, which avoids the singularity through a de Sitter core. Ay\'{o}n-Beato and Garc\'{ı}a obtained the first regular black hole in Einstein gravity by considering the coupling between Einstein gravity and a nonlinear electromagnetic field \cite{Ayon-Beato:1998hmi}. They later demonstrated that the Bardeen black hole solution can be rederived within this framework, showing that the Bardeen model can be interpreted as a nonlinear magnetic monopole in nonlinear electrodynamics \cite{Ayon-Beato:2000mjt}. Subsequently, numerous other regular black holes were proposed by incorporating various forms of nonlinear electrodynamics \cite{Hayward:2005gi,Berej:2006cc,Burinskii:2002pz,Balart:2014cga,Ma:2015gpa,Nam:2018uvc}. In 2016, Fan et al.\ presented a general method for obtaining exact charged (electric or magnetic) regular black hole solutions in general relativity coupled to nonlinear electrodynamics, and provided a class of regular black hole solutions \cite{Fan:2016hvf}. However, there is an inconsistency between the black hole solutions they presented and the corresponding Lagrangian densities, rendering their solutions ``incomplete.'' Toshmatov et al.\ provided the correct form of the mass function \cite{Toshmatov:2018cks}.

Although regular black holes represent a significant step towards resolving the singularity problem, their thermodynamic properties suffer from a fundamental puzzle. Based on the pioneering work of Bekenstein and Hawking, black holes are regarded as thermodynamic systems with temperature and entropy, obeying the laws of black hole thermodynamics. However, for many regular black holes, the geometrically defined Hawking temperature (calculated via surface gravity) is inconsistent with the temperature derived from the first law of thermodynamics (assuming the entropy satisfies the Bekenstein--Hawking area law $S = A/4$) \cite{Ma:2014qma}. This inconsistency signals more than a temperature problem---it indicates that previous thermodynamic studies of regular black holes are not self-consistent.

Recently, Ma et al.\ revealed the fundamental origin of this difficulty \cite{Ma:2025dee}. They pointed out that the construction of regular black holes usually introduces an extra constraint among the black hole parameters (such as the mass $M$, charge $Q$, coupling constant $\alpha$, etc.)---for instance, requiring the mass function to vanish at the origin to ensure regularity:
\begin{equation}\label{eq:11}
    \lim_{r \to 0} m(r) = 0 .
\end{equation}
This constraint not only reduces the dimension of the thermodynamic phase space, but, more importantly, modifies the first law of thermodynamics in that phase space.

To illustrate this point, we take the Reissner--Nordstr\"om (RN) black hole as an example. The first law of black hole thermodynamics for an RN black hole reads
\begin{equation}
dM = T\,dS + \Phi\,dQ. \label{eq:12}
\end{equation}
For an extremal RN black hole, which satisfies the condition $Q=M$, one might hastily rewrite the above equation as
\begin{equation}
dM = \frac{T}{1-\Phi}\,dS. \label{eq:13}
\end{equation}
If one computes the Hawking temperature from Eq.(\ref{eq:13}), one obtains a temperature $T^{(II)}=\frac{T}{1-\Phi}$, whereas the Hawking temperature $T_g$ calculated from the surface gravity gives a different result. The reason is that Eq.(\ref{eq:12}) is defined in the full phase space and the constraint $Q=M$ cannot be directly substituted into it. If one does so, not only does the Hawking temperature become incorrect, but the charge term disappears. The correct procedure is to first derive the expressions for the thermodynamic quantities from Eq.(\ref{eq:12}), and then substitute the condition $Q=M$ into those expressions. Let us take the calculation of the Hawking temperature for an extremal RN black hole as an example. From Eq.(\ref{eq:12}) we obtain
\begin{equation}
T_1 = \frac{\partial M}{\partial S} = \frac{r_+^2 - Q^2}{4\pi r_+^3}. \label{eq:14}
\end{equation}
Imposing the extremal condition $Q=M$ yields
\begin{equation}
T_2 = T_g=0 . \label{eq:15}
\end{equation}
In this way, one obtains the correct Hawking temperature of the extremal RN black hole via the thermodynamic method. Many previous studies on the thermodynamics of regular black holes were carried out by directly imposing the regularity constraint on the first law, which naturally leads to a discrepancy between the geometrically defined Hawking temperature and the temperature derived from the first law. This temperature mismatch, however, is only a symptom of a deeper issue. As the RN example shows, when the constraint is imposed prematurely, the charge term itself disappears from the first law. The problem is therefore not confined to the temperature---it indicates that the entire differential structure of the first law is distorted.

Real astrophysical black holes in the Universe are almost always rotating \cite{Mukhopadhyay:2018smw, Bambi}. Angular momentum not only significantly influences the geometric structure and dynamical processes of black holes, but is also an indispensable key parameter in any astrophysical observation \cite{Kumar:2020yem, Afrin:2021imp}. Motivated by this, we extend the idea of Ma et al.\ to the rotating case. Applying the modified Newman--Janis algorithm, we first construct the rotating ``mother'' black hole, whose parameters $M$, $q$, and $\alpha$ remain independent---analogous to the RN black hole in the example above. Imposing the regularity constraint $M=q^3/\alpha$ on it then yields the rotating regular black hole, which corresponds to the extremal RN black hole. We first derive all thermodynamic quantities in the unconstrained phase space, and only then impose the regularity constraint, thereby obtaining a self-consistent thermodynamic description. A key finding of our analysis is the asymmetry between the geometric and thermodynamic derivations: the geometric Hawking temperature is insensitive to the order in which the constraint is imposed, while the thermodynamic temperature is not---the correct result is obtained only when the constraint is imposed after the thermodynamic derivation in the full phase space. This ``full phase space first'' prescription resolves the inconsistency at the structural level, simultaneously guaranteeing the correctness of all thermodynamic quantities and thereby establishing a self-consistent thermodynamic framework for rotating regular black holes.

The structure of this paper is as follows. In Sec.~2, we review the regularity condition for a class of regular black holes and derive the corresponding spherically symmetric ``mother'' black hole. In Sec.~3, we apply the modified Newman--Janis algorithm to construct the rotating singular ``mother'' black hole. In Sec.~4, we first derive the thermodynamic quantities of the rotating ``mother'' black hole in the full phase space, and then impose the regularity constraint to obtain the correct thermodynamics of the rotating regular black hole. We contrast this with the incorrect result obtained when the constraint is imposed beforehand, and reveal the asymmetry between the geometric and thermodynamic derivations. Finally, in Sec.~5, we summarize the main conclusions.

{\it Constraint condition for regular black holes-}\label{sec:2} In this section, we present the spherically symmetric ``mother'' black hole solution and review how the regularity condition is obtained. The coupling between Einstein gravity and nonlinear electrodynamics is described by the following equations:
\begin{equation}\label{eq:21}
\begin{array}{c}
G_{\mu\nu} = T_{\mu\nu}, \quad \nabla_\mu (\mathcal{L}_{\mathcal{F}} F^{\mu\nu}) = 0,
\end{array}
\end{equation}
where $\mathcal{L}_{\mathcal{F}}=\partial\mathcal{L}/\partial\mathcal{F}$ is the derivative of the Lagrangian density $\mathcal{L}$ with respect to the invariant $\mathcal{F}=F_{\mu\nu}F^{\mu\nu}$. The energy–momentum tensor is
\begin{equation}\label{eq:22}
T_{\mu \nu} = 2 \left( \mathcal{L}_{\mathcal{F}} F_{\mu \nu}^2 - \frac{1}{4} g_{\mu \nu} \mathcal{L} \right).
\end{equation}

The static spherically symmetric black hole with nonlinear electric/magnetic charges has a line element given by
\begin{equation}\label{eq:23}
\begin{array}{c}
ds^2=-f(r) \mathrm{d} t^{2}+f(r)^{-1} \mathrm{~d} r^{2}+r^{2} \mathrm{~d} \Omega^{2}, \\
			\mathrm{~d} \Omega^{2}=d \theta^{2}+\sin ^{2} \theta d \phi^{2},\\
			\quad f(r)=1-\frac{2m(r)}{r},
\end{array}
\end{equation}
where $m(r)$ is the mass distribution function. Substituting the above equation into Eq.(\ref{eq:21}), the expression for the Lagrangian density is given by \cite{Fan:2016hvf}
\begin{equation}\label{eq:24}
\begin{array}{c}
\mathcal{L} = \frac{4 m'(r)}{r^2}.
\end{array}
\end{equation}
Integrating the above equation and considering the boundary condition $M=\lim\limits_{r \rightarrow \infty}m(r)$ leads to
\begin{equation}\label{eq:25}
\begin{array}{c}
m(r) = M -\frac{1}{4}\int_{r}^{\infty} r^{2} \mathcal{L}dr,
\end{array}
\end{equation}
where $M$ is the ADM mass in asymptotically flat spacetime. For $\mathcal{L}=0$ and $\mathcal{L}=\mathcal{F}$, one obtains $m(r)=$const and $m(r)=M-Q_m^2/(2r)$, corresponding to the metric functions of the Schwarzschild black hole and the Reissner-Nordstr\"{o}m black hole, respectively.

The expression for the Lagrangian density of a class of regular black holes is \cite{Fan:2016hvf}
\begin{equation}\label{eq:26}
\begin{array}{c}
\mathcal{L} = \frac{4\mu}{\alpha} \frac{\left(\alpha\mathcal{F}\right)^{\frac{\nu+3}{4}}}{\left( 1 + \left(\alpha\mathcal{F}\right)^{\frac{\nu}{4}}\right)^{\frac{\mu+\nu}{\nu}}},
\end{array}
\end{equation}
where $\mathcal{F} = 2Q^2_m/r^4$ with $Q_m = q^2/\sqrt{2\alpha}$, $q$ being the regularization parameter, while $\alpha>0$ has dimensions of length squared and $\mu > 0$ is a dimensionless constant. By substituting Eq.(\ref{eq:26}) into Eq.(\ref{eq:25}), one obtains \cite{Toshmatov:2018cks}
\begin{equation}\label{eq:27}
\begin{array}{c}
m(r) = M - \frac{q^3}{\alpha} \left[1 - \frac{r^{\mu}}{(r^\nu + q^\nu)^{\mu/\nu}}\right],
\end{array}
\end{equation}
where $\mu \ge 3$. Therefore, the metric function takes the form
\begin{equation}\label{eq:28}
\begin{array}{c}
f(r)=1-\frac{2}{r}\left(M-\frac{q^3}{\alpha} \left(1-\frac{ r^{\mu}}{(r^\nu + q^\nu)^{\frac{\mu}{\nu}}}\right)\right).
\end{array}
\end{equation}
The above equation is exactly the metric function of the ``mother'' black hole of a class of regular black holes.

Considering the regularity condition for regular black holes, substituting Eq.(\ref{eq:27}) into Eq.(\ref{eq:11}), one obtains \cite{Fan:2016hvf,Toshmatov:2018cks,Ma:2025dee}
\begin{equation}\label{eq:29}
\begin{array}{c}
M=\frac{q^3}{\alpha} .
\end{array}
\end{equation}
The above equation is the constraint condition for a class of regular black holes. When this constraint is substituted into the ``mother'' black hole metric, one obtains the corresponding regular black hole. Substituting Eq.(\ref{eq:29}) into Eq.(\ref{eq:28}), one obtains the metric function of a class of spherically symmetric regular black holes \cite{Toshmatov:2018cks}:
\begin{equation}\label{eq:30}
\begin{array}{c}
f(r) = 1 - \frac{2M r^{\mu-1}}{(r^\nu + q^\nu)^{\frac{\mu}{\nu}}},
\end{array}
\end{equation}
which reduces to the Bardeen black hole metric for $\mu = 3$, $\nu = 2$ and to the Hayward black hole metric for $\mu = \nu = 3$. For this spherically symmetric black hole solution, the Kretschmann scalar remains finite everywhere, and the solutions are regular \cite{Maluf:2018lyu}.

{\it The rotating singular ``mother'' black hole -}\label{sec:3}
In 1965, Newman and Janis first proposed the Newman--Janis algorithm (NJA), which derives the Kerr metric by performing complex coordinate transformations on the Schwarzschild line element \cite{Newman:1965tw}. Subsequently, the NJA has been widely adopted to generate rotating solutions from various static spherically symmetric black holes \cite{Kim:2019hfp,Benavides-Gallego:2018odl,Liu:2020ola,Xu:2020jpv,Kumar:2017qws,Xu:2016jod,Mazza:2021rgq}. In the context of regular black holes, Bambi and Modesto first utilized this algorithm to successfully construct the rotating metrics for Hayward and Bardeen black holes \cite{Bambi:2013ufa}. The resulting regular rotating metrics can be categorized into Type-I and Type-II, where Type-I solutions have a relatively concise form and can be expressed in a Kerr-like form within Boyer--Lindquist coordinates, and they have been extensively used as the rotating counterparts of regular black holes in subsequent studies \cite{Zhang:2020mxi,Beltracchi:2021ris,Toshmatov:2017zpr,Tsukamoto:2017fxq}.

However, the original NJA suffers from ambiguities in the complexification procedure, which can lead to the failure to bring the generated metric to Boyer--Lindquist coordinates. To overcome these difficulties, Azreg-A\"inou proposed a modified NJA that drops the complexification procedure entirely and instead determines the metric functions through physical requirements, successfully generating Type-I rotating solutions without the ambiguities of the original algorithm \cite{Azreg-Ainou:2014pra,Azreg-Ainou:2013ska}.

Applying the modified NJA \cite{Azreg-Ainou:2014pra,Azreg-Ainou:2013ska} to the spherically symmetric ``mother'' black hole, the line element of the rotating ``mother'' black hole is obtained as
\begin{equation}\label{eq:31}
    \begin{aligned}
       d s^{2}= & -\left[1-\frac{2 m(r) r}{\Sigma}\right] d t^{2}+\frac{\Sigma}{\Delta} d r^{2}-\frac{4 a m(r) r \sin ^{2} \theta}{\Sigma} d t d \phi \\
& +\Sigma d \theta^{2}+\sin ^{2} \theta\left[r^{2}+a^{2}+\frac{2 a^{2} m(r) r \sin ^{2} \theta}{\Sigma}\right] d \phi^{2},
    \end{aligned}
\end{equation}
where
\begin{equation}\label{eq:32}
    \begin{aligned}
      \Delta &= r^{2}-2 m(r) r+a^{2}, \\
      \Sigma &= r^{2}+a^{2} \cos ^{2} \theta, \\
      m(r) & = M - \frac{q^3}{\alpha} \left[1 - \frac{r^{\mu}}{(r^\nu + q^\nu)^{\mu/\nu}}\right].
    \end{aligned}
\end{equation}
The parameters $M$, $a$, $q$, and $\alpha$ in Eqs.(\ref{eq:31})--(\ref{eq:32}) are all independent, so the metric describes a rotating singular black hole. Imposing the regularity condition $M = q^3/\alpha$ on Eq.(\ref{eq:32}) yields the rotating regular black hole \cite{Kar:2025anc}.

{\it Thermodynamics of rotating regular black holes -}In the previous section, the rotating singular ``mother'' black hole was obtained, described by Eqs.(\ref{eq:31}) and (\ref{eq:32}), where the parameters $M$, $q$, and $\alpha$ are still independent. In this section, we first derive the thermodynamic quantities of the rotating ``mother'' black hole in the full phase space $\{S, J, q, \alpha\}$, and then impose the regularity condition $M = q^3/\alpha$ on the derived quantities to obtain the thermodynamics of the rotating regular black hole. For comparison, we also examine the incorrect approach, in which the constraint is imposed directly on the first law.

The event horizon corresponds to the Killing horizon, which defines the ``surface'' of a black hole as a three-dimensional hypersurface. For the rotating ``mother'' black hole described by the line element Eq.(\ref{eq:31}), the event horizon satisfies
\begin{equation}\label{eq:41}
r_H^2+a^2 -2r_H\left(M - \frac{q^3}{\alpha} \left(1 - \frac{r_H^{\mu}}{(r_H^\nu + q^\nu)^{\mu/\nu}}\right)\right)= 0,
\end{equation}
where $r_H$ is the event horizon radius.

For the rotating ``mother'' black hole, the independent thermodynamic extensive quantities include: the entropy $S$, the angular momentum $J$, the regularization parameter $q$, and the parameter $\alpha$. In this phase space $\{S,\ J,\ q,\ \alpha\}$, from Eq.(\ref{eq:41}), we obtain the squared-mass formula as
\begin{equation}\label{eq:42}
    \begin{aligned}
     M^2 = &\frac{\pi}{4S} \left[\frac{2 q^3}{\alpha} \left(1-\frac{\left(\frac{S}{\pi }-a^2\right)^{\frac{\mu}{2}}}{\left(\left(\frac{S}{\pi }-a^2\right)^{\frac{\nu}{2}}+q^{\nu }\right)^{\mu /\nu }}\right)\right.\\
    &\left.\left(\frac{S}{\pi }-a^2\right)^{\frac{1}{2}}+\frac{S}{\pi }\right]^2+\frac{\pi }{S} J^2,
    \end{aligned}
\end{equation}
where $S=\pi(r_H^2+a^2)$ and $J=Ma$. In this phase space, the first law of thermodynamics is
\begin{equation}
dM = T dS + \Omega dJ + \Psi dq + \mathcal{A} d\alpha,
\label{eq:43}
\end{equation}
where
\begin{equation}\label{eq:44}
    \begin{aligned}
T &\equiv \left( \frac{\partial M}{\partial S} \right)_{J, q, \alpha}, \quad
\Omega \equiv \left( \frac{\partial M}{\partial J} \right)_{S, q, \alpha}, \\
\Psi &\equiv \left( \frac{\partial M}{\partial q} \right)_{S, J, \alpha}, \quad
\mathcal{A} \equiv \left( \frac{\partial M}{\partial \alpha} \right)_{S, J, q}.
    \end{aligned}
\end{equation}
From Eqs.~(\ref{eq:42}) and (\ref{eq:44}), one can compute all the thermodynamic quantities of the rotating ``mother'' black hole in the full phase space.

With the full phase space established, we derive the Hawking temperature of the rotating ``mother'' black hole. From Eqs.(\ref{eq:42}) and (\ref{eq:44}), we obtain
\begin{equation}\label{eq:45}
    \begin{aligned}
T_1^{(I)} =&\frac{r q^{\nu }-\frac{2 \mu}{\alpha}   q^{\nu +3} r^{\mu } \left(q^{\nu }+r^{\nu }\right)^{-\frac{\mu }{\nu }}+r^{\nu +1}}{4 \pi  \left(a^2+r^2\right)(q^{\nu }+r^{\nu })}\\
&-\frac{a^2}{4 \pi r  \left(a^2+r^2\right)}.
    \end{aligned}
\end{equation}
To verify this result, we compare it with the geometric definition. At the event horizon $r_H$, the Hawking temperature is proportional to the surface gravity $\kappa$:
\begin{equation}\label{eq:46}
    \begin{aligned}
T = \frac{\kappa}{2\pi} = \frac{\partial_{r_H} \Delta(r_H)}{4\pi (r_H^2 + a^2)}.
    \end{aligned}
\end{equation}
Substituting Eq.(\ref{eq:41}) into Eq.(\ref{eq:46}) to calculate $T_{1g}$, we obtain $T_{1g} = T_1^{(I)}$. This confirms the consistency between the geometric and thermodynamic descriptions of the rotating ``mother'' black hole in the full phase space.

Substituting the regularity constraint $M = q^3/\alpha$ from Eq.(\ref{eq:29}) into $T_1^{(I)}$ and $T_{1g}$ yields 
\begin{equation}\label{eq:47}
    \begin{aligned}
T_2^{(I)} &= T_{2g}\\
&=\frac{r^2 \left(r^{\nu }-(\mu -1) q^{\nu }\right)-a^2 \left((\mu +1) q^{\nu }+r^{\nu }\right)}{4 \pi  r \left(a^2+r^2\right) \left(q^{\nu }+r^{\nu }\right)},
    \end{aligned}
\end{equation}
which is the physically correct Hawking temperature of the rotating regular black hole. Imposing the regularity condition on the remaining thermodynamic quantities derived from Eq.~(\ref{eq:44}) similarly yields their correct expressions for the rotating regular black hole. The equality $T_2^{(I)} = T_{2g}$ demonstrates that when the constraint is imposed after deriving the temperature in the full phase space, the thermodynamic result coincides with the geometric one. Since the first law Eq.~(\ref{eq:43}) was established in the full phase space and all thermodynamic quantities were derived before the constraint was imposed, the temperature equality is not an isolated fix---it serves as a nontrivial verification that the entire set of thermodynamic quantities is internally consistent and physically correct.

We now contrast this with the incorrect alternative---imposing the constraint on the first law of thermodynamics beforehand. Substituting the constraint $M = q^3/\alpha$ from Eq.(\ref{eq:29}) directly into the first law Eq.(\ref{eq:43}), one obtains
\begin{equation}\label{eq:411}
    \begin{aligned}
dM = \frac{1}{1 + \dfrac{\mathcal{A}_2 q^3}{M^2}} \Bigg[ T_2^{(I)}\, dS + \Omega_2\, dJ + \left( \Psi_2 + \frac{3\mathcal{A}_2 q^2}{M} \right) dq \Bigg],
    \end{aligned}
\end{equation}
where $M = \sqrt{ \frac{S}{4\pi} \left( \frac{ \left( \left(\frac{S}{\pi} - a^2\right)^{\frac{\nu}{2}} + q^{\nu} \right)^{\mu/\nu}}{\left( \frac{S}{\pi} - a^2 \right)^{\mu/2}} \right)^2 + \frac{\pi }{S} J^2}$, and $\Omega_2$, $\Psi_2$, $\mathcal{A}_2$ are the conjugate quantities evaluated under the constraint $M = q^3/\alpha$. From this distorted differential relation, the resulting thermodynamic temperature is
\begin{equation}\label{eq:49}
    \begin{aligned}
T^{(II)}&=\frac{r^2 \left(r^\nu + (1-\mu) q^\nu\right) - a^2 \left(r^\nu + (1+\mu) q^\nu\right)}{4\pi r^{1+\mu} (a^2 + r^2) (q^\nu + r^\nu)^{1 - \frac{\mu}{\nu}}}.
    \end{aligned}
\end{equation}
This $T^{(II)}$ is merely an effective temperature, related to the physical temperature by $T^{(II)}=\frac{T_2^{(I)}M^2}{M^2 + \mathcal{A}_2 q^3}$, and lacks direct physical meaning.

In contrast, the geometric method is insensitive to the order of constraint application. Imposing the constraint condition Eq.(\ref{eq:29}) on the ``mother'' black hole metric yields the rotating regular black hole. The event horizon of this regular black hole then satisfies
\begin{equation}\label{eq:48}
    \begin{aligned}
\Delta=r_H^{2}-2 \frac{M r_H^{\mu+1}}{(r_H^\nu + q^\nu)^{\frac{\mu}{\nu}}}+a^{2}=0.
    \end{aligned}
\end{equation}
Using the geometric definition of Hawking temperature Eq.(\ref{eq:46}) together with this horizon equation yields
\begin{equation}\label{eq:410}
    \begin{aligned}
T_g&=T_{2g}\\
   &=\frac{r^2 \left(r^{\nu }-(\mu -1) q^{\nu }\right)-a^2 \left((\mu +1) q^{\nu }+r^{\nu }\right)}{4 \pi  r \left(a^2+r^2\right) \left(q^{\nu }+r^{\nu }\right)},
    \end{aligned}
\end{equation}
which coincides with $T_2^{(I)}$, confirming that the geometric result is always physically correct regardless of the order of constraint imposition.

The above analysis reveals a fundamental asymmetry between the geometric and thermodynamic derivations. From the geometric perspective, the Hawking temperature is invariant under the order of constraint application: whether one evaluates the surface gravity using the unconstrained metric and then imposes $M=q^3/\alpha$, or first imposes the constraint on the metric and then computes the surface gravity, the result is always $T_{2g}$. Mathematically, this follows from the chain rule, since the surface gravity involves only derivatives of the metric functions evaluated at the horizon---the geometric temperature is an intrinsic property of the spacetime.

In the thermodynamic approach, by contrast, the order of operations is critical. When the constraint $M=q^3/\alpha$ is used to eliminate $\alpha$ in the first law Eq.(\ref{eq:43}), the resulting differential relation acquires a nontrivial prefactor [see Eq.(\ref{eq:411})], and the resulting $T^{(II)}$ does not equal the physical temperature $T_2^{(I)}$. The correct procedure is therefore to compute all thermodynamic quantities in the unconstrained phase space $\{S,J,q,\alpha\}$, and only afterwards impose the regularity condition. This preserves the proper differential structure of the first law and ensures that the thermodynamic temperature coincides with its geometric counterpart.

When the constraint is imposed on the first law before the thermodynamic derivation, one obtains the distorted relation Eq.(\ref{eq:411}), it introduces a common prefactor $1/(1+\mathcal{A}_2 q^3/M^2)$ that multiplies every term in the first law, thereby distorting all thermodynamic quantities. In other words, the error is not confined to the temperature: all thermodynamic quantities are rendered effective rather than physical. Previous thermodynamic analyses of regular black holes that started from the constrained phase space therefore did not possess a self-consistent thermodynamic description at all. The ``full phase space first'' prescription resolves this at the structural level, simultaneously guaranteeing the correctness of all thermodynamic quantities.

{\it Conclusion -}
In this study, we have established a self-consistent thermodynamic framework for rotating regular black holes. By constructing the rotating ``mother'' black hole with all parameters independent, and systematically analyzing the order in which the regularity constraint is imposed, we arrive at the following conclusions:

(1) The geometric Hawking temperature is insensitive to the order in which the constraint $M = q^3/\alpha$ is applied. One may either first compute the surface gravity using the unconstrained ``mother'' black hole metric and then substitute $M = q^3/\alpha$ into the resulting expression, or first substitute the constraint into the metric and then evaluate the surface gravity; both procedures yield the same temperature $T_{2g}=T_g$. Mathematically, this follows from the chain rule: the surface gravity involves only derivatives of the metric functions at the horizon, so the two procedures give the same result.

(2) The thermodynamic temperature, by contrast, is sensitive to the order of operations. If one eliminates $\alpha$ in the first law Eq.(\ref{eq:43}) using the constraint $M = q^3/\alpha$ before deriving the temperature, the resulting differential relation acquires a nontrivial prefactor [see Eq.(\ref{eq:411})], and the temperature obtained, $T^{(II)}$, is merely an effective temperature without direct physical meaning. The correct procedure is to first derive the temperature from the first law in the unconstrained phase space $\{S, J, q, \alpha\}$, and only afterwards impose the regularity condition. This yields the physical temperature $T_2^{(I)} = T_{2g}$, consistent with the geometric result.

(3) The temperature equality $T_2^{(I)} = T_{2g}$ is not merely a resolution of a calculational inconsistency---it is a verification that the thermodynamic framework established through the ``full phase space first'' prescription is fully self-consistent. When the constraint is incorrectly imposed on the first law beforehand, not only the temperature but all thermodynamic quantities are distorted by the same nontrivial prefactor, rendering the entire thermodynamic description unreliable. By working in the unconstrained phase space first and imposing the constraint only on the derived quantities, we obtain correct expressions for all thermodynamic quantities simultaneously.

The significance of this work lies not only in resolving the temperature discrepancy, but in establishing a self-consistent thermodynamic framework for rotating regular black holes. The equality $T_2^{(I)} = T_{2g}$ between the thermodynamic and geometric temperatures is the signature of this self-consistency, not its goal. With the internal consistency of the thermodynamic description now assured, this opens the way for a systematic investigation of phase transitions, critical phenomena, thermodynamic stability, and microscopic degrees of freedom.

{\it Acknowledgments-}
The work is supported by Yunnan Xingdian Talent Support Program - Young Talent Project, and the National Natural Science Foundation of China (Grant No. 12463001).


\begin{thebibliography}{99}

\bibitem{Hawking:1970zqf}
S.W. Hawking and R. Penrose, Proc. Roy. Soc. Lond. A \textbf{314} (1970) 529-548.

\bibitem{bardeen1968non}
J.~Bardeen,
GR5, 87 (1968).

\bibitem{Ayon-Beato:1998hmi}
E.~Ayon-Beato and A.~Garcia,
Phys. Rev. Lett. \textbf{80} (1998), 5056-5059
doi:10.1103/PhysRevLett.80.5056
[arXiv:gr-qc/9911046 [gr-qc]].

\bibitem{Ayon-Beato:2000mjt}
E.~Ayon-Beato and A.~Garcia,
Phys. Lett. B \textbf{493} (2000), 149-152
doi:10.1016/S0370-2693(00)01125-4
[arXiv:gr-qc/0009077 [gr-qc]].


\bibitem{Hayward:2005gi}
S.~A.~Hayward,
Phys. Rev. Lett. \textbf{96} (2006), 031103
doi:10.1103/PhysRevLett.96.031103
[arXiv:gr-qc/0506126 [gr-qc]].

\bibitem{Berej:2006cc}
W.~Berej, J.~Matyjasek, D.~Tryniecki and M.~Woronowicz,
Gen. Rel. Grav. \textbf{38} (2006), 885-906
doi:10.1007/s10714-006-0270-9
[arXiv:hep-th/0606185 [hep-th]].

\bibitem{Burinskii:2002pz}
A.~Burinskii and S.~R.~Hildebrandt,
Phys. Rev. D \textbf{65} (2002), 104017
doi:10.1103/PhysRevD.65.104017
[arXiv:hep-th/0202066 [hep-th]].

\bibitem{Balart:2014cga}
L.~Balart and E.~C.~Vagenas,
Phys. Rev. D \textbf{90} (2014) no.12, 124045
doi:10.1103/PhysRevD.90.124045
[arXiv:1408.0306 [gr-qc]].

\bibitem{Ma:2015gpa}
M.~S.~Ma,
Annals Phys. \textbf{362} (2015), 529-537
doi:10.1016/j.aop.2015.08.028
[arXiv:1509.05580 [gr-qc]].

\bibitem{Nam:2018uvc}
C.~H.~Nam,
Gen. Rel. Grav. \textbf{50} (2018) no.6, 57
doi:10.1007/s10714-018-2380-6

\bibitem{Fan:2016hvf}
Z.~Y.~Fan and X.~Wang,
Phys. Rev. D \textbf{94} (2016) no.12, 124027
doi:10.1103/PhysRevD.94.124027
[arXiv:1610.02636 [gr-qc]].

\bibitem{Toshmatov:2018cks}
B.~Toshmatov, Z.~Stuchl{\'\i}k and B.~Ahmedov,
Phys. Rev. D \textbf{98} (2018) no.2, 028501
doi:10.1103/PhysRevD.98.028501
[arXiv:1807.09502 [gr-qc]].

\bibitem{Ma:2014qma}
M.~S.~Ma and R.~Zhao,
Class. Quant. Grav. \textbf{31} (2014), 245014
doi:10.1088/0264-9381/31/24/245014
[arXiv:1411.0833 [gr-qc]].

\bibitem{Ma:2025dee}
M.~S.~Ma, H.~F.~Li and J.~H.~Shi,
Sci. China Phys. Mech. Astron. \textbf{69} (2026) no.1, 210411
doi:10.1007/s11433-025-2753-6
[arXiv:2507.09551 [gr-qc]].

\bibitem{Mukhopadhyay:2018smw}
B.~Mukhopadhyay,
Astrophys. Space Sci. Proc. \textbf{53} (2018), 3-15
doi:10.1007/978-3-319-94607-8{\_}1

\bibitem{Bambi}
C, Bambi, Astrophysical Black Holes: A Review, arXiv.1906.03871 [gr-qc].

\bibitem{Kumar:2020yem}
R.~Kumar, A.~Kumar and S.~G.~Ghosh,
Astrophys. J. \textbf{896} (2020) no.1, 89
doi:10.3847/1538-4357/ab8c4a
[arXiv:2006.09869 [gr-qc]].

\bibitem{Afrin:2021imp}
M.~Afrin, R.~Kumar and S.~G.~Ghosh,
Mon. Not. Roy. Astron. Soc. \textbf{504} (2021) no.4, 5927-5940
doi:10.1093/mnras/stab1260
[arXiv:2103.11417 [gr-qc]].

\bibitem{Maluf:2018lyu}
R.~V.~Maluf and J.~C.~S.~Neves,
Phys. Rev. D \textbf{97} (2018) no.10, 104015
doi:10.1103/PhysRevD.97.104015
[arXiv:1801.02661 [gr-qc]].


\bibitem{Newman:1965tw}
E.~T.~Newman and A.~I.~Janis,
J. Math. Phys. \textbf{6} (1965), 915-917
doi:10.1063/1.1704350

\bibitem{Kim:2019hfp}
H.~C.~Kim, B.~H.~Lee, W.~Lee and Y.~Lee,
Phys. Rev. D \textbf{101} (2020) no.6, 064067
doi:10.1103/PhysRevD.101.064067
[arXiv:1912.09709 [gr-qc]].

\bibitem{Benavides-Gallego:2018odl}
C.~A.~Benavides-Gallego, A.~A.~Abdujabbarov and C.~Bambi,
Phys. Rev. D \textbf{101} (2020) no.4, 044038
doi:10.1103/PhysRevD.101.044038
[arXiv:1811.01562 [gr-qc]].

\bibitem{Liu:2020ola}
C.~Liu, T.~Zhu, Q.~Wu, K.~Jusufi, M.~Jamil, M.~Azreg-A{\"\i}nou and A.~Wang,
Phys. Rev. D \textbf{101} (2020) no.8, 084001
[erratum: Phys. Rev. D \textbf{103} (2021) no.8, 089902]
doi:10.1103/PhysRevD.101.084001
[arXiv:2003.00477 [gr-qc]].

\bibitem{Xu:2020jpv}
Z.~Xu, X.~Gong and S.~N.~Zhang,
Phys. Rev. D \textbf{101} (2020) no.2, 024029
doi:10.1103/PhysRevD.101.024029

\bibitem{Kumar:2017qws}
R.~Kumar and S.~G.~Ghosh,
Eur. Phys. J. C \textbf{78} (2018) no.9, 750
doi:10.1140/epjc/s10052-018-6206-1
[arXiv:1711.08256 [gr-qc]].

\bibitem{Xu:2016jod}
Z.~Xu and J.~Wang,
Phys. Rev. D \textbf{95} (2017) no.6, 064015
doi:10.1103/PhysRevD.95.064015
[arXiv:1609.02045 [gr-qc]].

\bibitem{Mazza:2021rgq}
J.~Mazza, E.~Franzin and S.~Liberati,
JCAP \textbf{04} (2021), 082
doi:10.1088/1475-7516/2021/04/082
[arXiv:2102.01105 [gr-qc]].


\bibitem{Bambi:2013ufa}
C.~Bambi and L.~Modesto,
Phys. Lett. B \textbf{721} (2013), 329-334
doi:10.1016/j.physletb.2013.03.025
[arXiv:1302.6075 [gr-qc]].

\bibitem{Beltracchi:2021ris}
P.~Beltracchi and P.~Gondolo,
Phys. Rev. D \textbf{104} (2021) no.12, 124066
doi:10.1103/PhysRevD.104.124066
[arXiv:2104.02255 [gr-qc]].

\bibitem{Toshmatov:2017zpr}
B.~Toshmatov, Z.~Stuchl{\'\i}k and B.~Ahmedov,
Phys. Rev. D \textbf{95} (2017) no.8, 084037
doi:10.1103/PhysRevD.95.084037
[arXiv:1704.07300 [gr-qc]].

\bibitem{Tsukamoto:2017fxq}
N.~Tsukamoto,
Phys. Rev. D \textbf{97} (2018) no.6, 064021
doi:10.1103/PhysRevD.97.064021
[arXiv:1708.07427 [gr-qc]].

\bibitem{Zhang:2020mxi}
H.~X.~Zhang, Y.~Chen, T.~C.~Ma, P.~Z.~He and J.~B.~Deng,
Chin. Phys. C \textbf{45} (2021) no.5, 055103
doi:10.1088/1674-1137/abe84c
[arXiv:2007.09408 [gr-qc]].

\bibitem{Azreg-Ainou:2014pra}
M.~Azreg-A{\"\i}nou,
Phys. Rev. D \textbf{90} (2014) no.6, 064041
doi:10.1103/PhysRevD.90.064041
[arXiv:1405.2569 [gr-qc]].

\bibitem{Azreg-Ainou:2013ska}
M.~Azreg-Ainou,
Eur. Phys. J. C \textbf{74} (2014), 2930
doi:10.1140/epjc/s10052-014-2930-3
[arXiv:1311.6963 [gr-qc]].



\bibitem{Kar:2025anc}
A.~Kar, A.~Dey and S.~Kar,
[arXiv:2510.11364 [gr-qc]].



\end{thebibliography}
\end{document}